\documentclass[sigconf]{acmart}

\usepackage{booktabs} 
\usepackage{algorithm}
\usepackage[noend]{algorithmic}
\usepackage{graphicx}
\usepackage{url}

\setcopyright{none}

\renewcommand\footnotetextcopyrightpermission[1]{}
\begin{document}

\title{Query-Driven Knowledge Base Completion using Multimodal Path Fusion over Multimodal Knowledge Graph}

\author{Yang Peng}
\email{yangpengsnf@ufl.edu}
\affiliation{University of Florida}

\author{Daisy Zhe Wang}
\email{daisyw@ufl.edu}
\affiliation{University of Florida}

\begin{abstract}

Over the past few years, large knowledge bases have been constructed to store
massive amounts of knowledge. However, these knowledge bases are highly
incomplete, for example, over 70\% of people in Freebase have no known place
of birth.
To solve this problem, we propose a query-driven knowledge base completion
system with multimodal fusion of unstructured and structured information.

To effectively fuse unstructured information from the Web and structured information in
knowledge bases to achieve good performance, our system builds multimodal
knowledge graphs based on question answering and rule inference.
We propose a multimodal path fusion algorithm to rank candidate answers based on
different paths in the multimodal knowledge graphs, achieving much better performance
than question answering, rule inference and a baseline fusion algorithm.

To improve system efficiency, query-driven techniques are utilized to reduce
the runtime of our system, providing fast responses to user queries.
Extensive experiments have been conducted to demonstrate the effectiveness and
efficiency of our system.

\end{abstract}


\maketitle

\section{Introduction}

A knowledge base (KB) is usually a data store of structured information about
entities, relations and facts.
In recent years, huge knowledge bases, such as Freebase \cite{bollacker2008freebase},
NELL \cite{carlson2010toward}, YAGO \cite{suchanek2007yago} and DeepDive \cite{zhang2015deepdive},
have been constructed to host massive amounts of knowledge acquired from real-world datasets.
Despite their huge size, these knowledge bases are greatly incomplete.
For example, Freebase \cite{bollacker2008freebase} contains over 112 million
entities and 388 million facts, while over 70\% of people included in Freebase
have no known place of birth.
Therefore, knowledge base completion has drawn a lot of attention from researchers.

Formally speaking, knowledge base completion (KBC) is the task to fill in the
gaps in knowledge bases.
Facts inside KBs are usually represented in triplet format as
\textit{\textless subject, relation, object\textgreater}, for example
\textit{\textless Marvin\_Minsky\footnote{Marvin Lee Minsky
(August 9, 1927 - January 24, 2016) was an American cognitive scientist concerned
largely with research of artificial intelligence (AI), co-founder of the Massachusetts
Institute of Technology's AI laboratory, and author of several texts concerning
AI and philosophy. \url{https://en.wikipedia.org/wiki/Marvin_Minsky}.},
wasBornIn, New\_York\_City\textgreater}.
A knowledge base completion query can be formulated as \textit{\textless subject, relation, ?\textgreater},
which means given the subject and relation, what is the corresponding object value(s)?

One popular way to fill in missing facts in a knowledge base is to extract
new facts from a large number of documents in batch mode.
Most knowledge bases are constructed using this approach iteratively on large
datasets (usually text corpora) \cite{bollacker2008freebase,carlson2010toward,suchanek2007yago}.
However, iterative batch processing is very time-consuming and requires intense
human labor involvement in preparing large datasets.
In contrast, we focus on a query-driven approach to knowledge base completion in this paper.
A user can issue a KBC query to our system and then our system searches the Web and
knowledge base to look for the missing facts to this query.
The query-driven paradigm enables a targeted and on-demand method for knowledge base completion.
And it's very flexible and fast compared to batch processing.

Previous approaches for query-driven knowledge base completion usually either
utilize only the unstructured textual information or only the structured information
inside knowledge bases.
However, structured information, such as entity types and entity-to-entity relatedness,
can help fact extraction tasks on unstructured datasets to achieve better quality.
And fusing unstructured and structured information together can further help improve performance,
because they have complementary strengths and weaknesses \cite{peng2016multimodal}.
In this paper, we propose a novel knowledge base completion system combining rule inference
and question answering and fusing unstructured text and structured knowledge bases.

In this paper, we design a multimodal path fusion algorithm to combine question answering
and rule inference in one unified framework.
Our system first generates a multimodal knowledge graph for each KBC query.
Then we use the multimodal path fusion algorithm to rank candidate answers based on
different paths between query entities and candidate answers.
The question answering system (WebQA) we implemented is explained in our previous paper \cite{arxivqa,peng2023web}.

We employ a few query-driven techniques to eliminate unnecessary computations
and reduce the runtime of our system pipeline.
We use query-driven optimization to avoid unnecessary WebQA operations in
multimodal knowledge graph generation, such as using confidence thresholds to
filter out unreliable results from WebQA.

Our contributions are shown below:
\begin{itemize}
	\item We propose an effective and efficient KBC system by combining rule inference
	and web-based question answering with the massive information available on the Web
	and in the knowledge bases.
	Our system fuses both unstructured data from the Web and structured information
	from knowledge bases in depth.
	\item 
	We generate a multimodal knowledge graph for each query based on question answering
	and rule inference.
	We propose a multimodal path fusion algorithm to effectively rank candidate answers
	using the multimodal knowledge graph.
	\item 
	To improve efficiency, we employ query-driven techniques for multimodal knowledge graph 
   generation to reduce the runtime and provide fast responses to user queries.
	\item 
	Extensive experiments have been conducted to demonstrate the effectiveness and
	efficiency of our system.
\end{itemize}

\noindent \textbf{Overview} Related work on knowledge base completion and multimodal fusion
is introduced and discussed in Section 2.
Our system and multimodal path fusion algorithm are presented in Section 3.
We demonstrate the effectiveness and efficiency of our system through extensive experiments
in Section 4.
The conclusions and future work of our KBC system are discussed in Section 5.

\section{Related Work}

There are several different types of approaches to fill in missing information
for incomplete knowledge bases.
In this section, we briefly discuss related work on batch-oriented knowledge base
construction, inference and learning inside knowledge bases and multimodal fusion.

\subsection{Batch-Oriented Knowledge Base Construction}

Huge knowledge bases have been constructed since mid-2000s \cite{bollacker2008freebase,carlson2010toward,suchanek2007yago,zhang2015deepdive}.
Most knowledge bases use iterative construction processes to learn extractors,
rules and facts from large datasets \cite{carlson2010toward,suchanek2007yago}.
Some knowledge bases employ human workers to manually add new information \cite{bollacker2008freebase}.

TAC KBP \cite{ji2010overview} and TREC KBA \cite{frank2012building} are the two
most famous annual competitions for knowledge base construction.
The goal of them is to develop and evaluate technologies for building and populating
knowledge bases from unstructured text.
Most of these methods process each document in turn, extracting as many facts
as possible by using named-entity linkage and (supervised) relation extraction methods.
Summaries of the standard approaches in TAC KBP and TREC KBA are given by
Ji and Grishman \cite{ji2011knowledge}, Weikum and Theobald \cite{weikum2010information} and Frank et al. \cite{frank2013evaluating}.

Manual KB construction is very slow and suffers from intense human labor cost.
Automatic knowledge base construction processes usually take very long time to finish,
e.g. a fast streaming system by Nia et al. \cite{niauniversity,nia2014streaming}
for TREC KBA still needs hours to process 5TB text data in one pass.
The low efficiency issue is even worse when people try to iteratively learn new extractors,
rules and facts on existing large datasets \cite{suchanek2007yago}
or create new datasets for knowledge base construction.
Another disadvantage of knowledge base construction is it can not guarantee to
extract the missing facts users are looking for in a targeted way.

\subsection{Inference and Learning}

Inference and statistical learning have been utilized in recent years for
knowledge base completion \cite{lao2011random,nickel2011three,richardson2006markov,tong2006fast}.
These approaches are usually restricted to the information only available inside knowledge bases.
To make the problem more difficult, information in knowledge bases are highly incomplete.

Logical rule inference has been widely used for inferring new facts in knowledge
bases \cite{chen2016ontological}.
Logical rules were first learned offline and then applied to infer new facts on-the-fly
in \cite{chen2016ontological}.
Richardson and Domingos proposed the Markov Logic Networks \cite{richardson2006markov}
for probabilistic inference based on logical rules and graphical models.
However, batch inference in Markov Logic Networks is very time-consuming and
unscalable for large knowledge bases.

Information inside knowledge bases can be structured as massive graphs of entities and relations.
Entities are treated as nodes and relations are treated as edges in the knowledge graphs.
Random walks in knowledge graphs have been utilized for knowledge base completion
for its scalability \cite{tong2006fast,lao2011random}.
Recent work \cite{nickel2011three,bordes2013translating,socher2013reasoning} learned
embedded representations of entities and relations in the knowledge bases and
used these representations to infer missing facts.
But learning expressive, scalable and effective models can be challenging
\cite{lao2011random,nickel2011three}.

\subsection{Multimodal Fusion}

\begin{figure*}[t]
	\centering
	\includegraphics[scale=0.9]{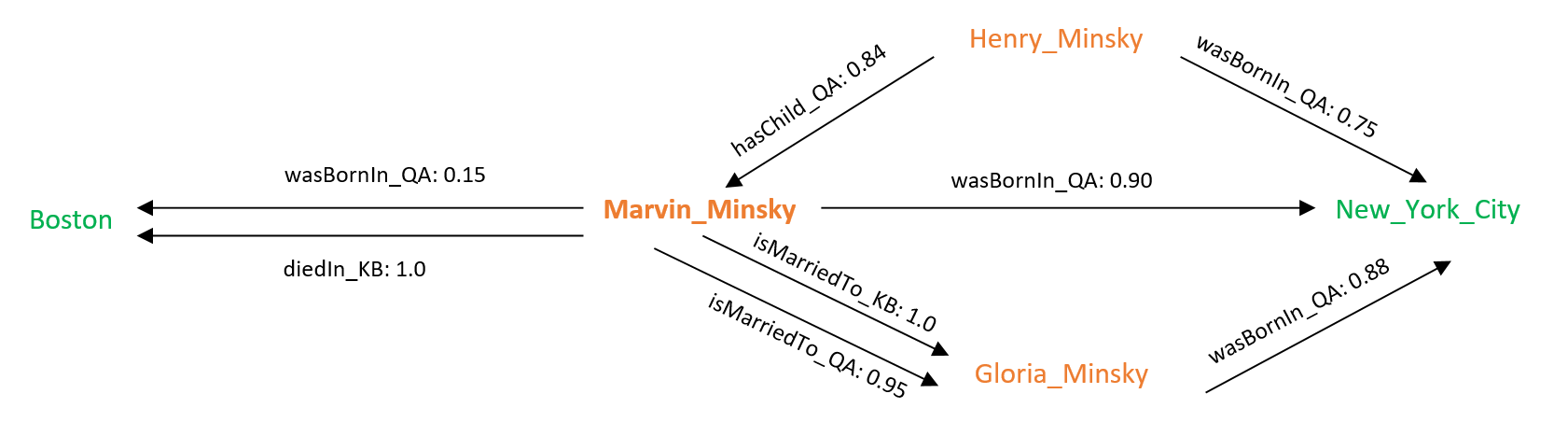}
	\caption{A simplified example of the multimodal knowledge graph.} \label{graph}
\end{figure*}

Multimodal fusion techniques have been widely used for tasks such as information retrieval, information extraction and classification tasks \cite{peng2015probabilistic, peng2016scalable, peng2016multimodal, peng2017multimodal, acm2017web}.
Multimodal fusion can utilize information from multiple types of data sources. 
By leveraging the complementary and correlative relations between different modalities \cite{peng2016multimodal}, multimodal fusion approaches can achieve higher performance than any single modality approaches in most cases.
In our previous work \cite{arxivqa,peng2023web}, we have utilized both unstructured textual information and structured information from knowledge bases to build a better question answering system for knowledge base completion.
In this paper, we take one step further to fuse both rule inference and web-based question answering to build a more capable knowledge base completion system.

\section{System}

To solve knowledge base completion, we propose a query-driven knowledge base completion
system with multimodal fusion of unstructured text and structured knowledge.
Although question answering and rule inference are utilized in our system, these two approaches
suffer from a few drawbacks.
Rule inference usually has low quality because it only uses existing facts
in knowledge bases which are highly incomplete.
Since question answering depends on search engines to extract missing information from the Web,
QA works poorly for rare entities on the Web.
So we design a new fusion method using multimodal knowledge graphs to combine unstructured
and structured information and overcome these problems.

Our system creates a multimodal knowledge graph for each KBC query by combining logical rules,
existing facts in knowledge bases and facts extracted from the Web.
Then we use the multimodal path fusion algorithm to rank candidate answers using paths in the
multimodal knowledge graph.
The intuition behind the multimodal path fusion algorithm is,
by linking query entities and candidate answers with semantic and expressive paths
generated from the Web and knowledge bases, we can effectively rank candidate answers based on
semantic information of these multimodal paths.

Since knowledge base completion is a never-ending process, providing fast responses
at query time is essential for users to quickly find out missing facts of interest.
Our system is a query-driven KBC system which uses query-driven techniques to avoid
unnecessary operations on-the-fly to improve efficiency, while previous work are
focused on batch-oriented systems.

\subsection{Multimodal Knowledge Graph}

We first introduce basic concepts of the multimodal knowledge graph (MKG).
A MKG is a directed graph consisting of nodes and edges, where nodes are entities
in knowledge bases and edges are relations between entities.
Subject entities, object entities and relations together form facts.
Different from traditional knowledge graphs which only include existing facts in
structured knowledge bases, the MKG contains facts extracted from the unstructured
Web using our WebQA \cite{arxivqa,peng2023web} system at query time.
So we have two types of edges in a MKG, one is directly loaded from knowledge bases,
the other is extracted from the Web at query time.
For example, for relation wasBornIn, there are two types of edges in Figure~\ref{graph},
\textit{wasBornIn\_KB} and \textit{wasBornIn\_QA}.

A MKG is path-centered, meaning the multimodal knowledge graph is generated according to paths.
A path is composed of a series of edges from query entities to candidate answers.
In our system, we mostly use horn-clause logical rules as paths.
For example, \textit{\textless hasChild\_QA$^-$, wasBornIn\_QA\textgreater} and
\textit{\textless isMarriedTo\_KB, wasBornIn\_KB\textgreater} are two paths based
on logical rules shown in Figure~\ref{graph}.
\textit{hasChild\_QA$^-$} is the reverse of relation \textit{hasChild\_QA}, for example,
a fact \textit{\textless Marvin\_Minksy, hasChild\_QA$^-$, Henry\_Minsky\textgreater} means
\textit{Henry\_Minksy} has a child \textit{Marvin\_Minsky}.
Note, there is a special type of paths with only one edge for each relation,
e.g. a path with only one edge \textit{wasBornIn\_QA} for relation \textit{wasBornIn}
in Figure~\ref{graph}, indicating the candidate answer is directly extracted from the Web.
These paths can be transformed into features to rank candidate answers, as shown
in Section 3.3.

\subsubsection{Logical Rules}

\begin{figure}[h]
	\centering
	\includegraphics[scale=0.5]{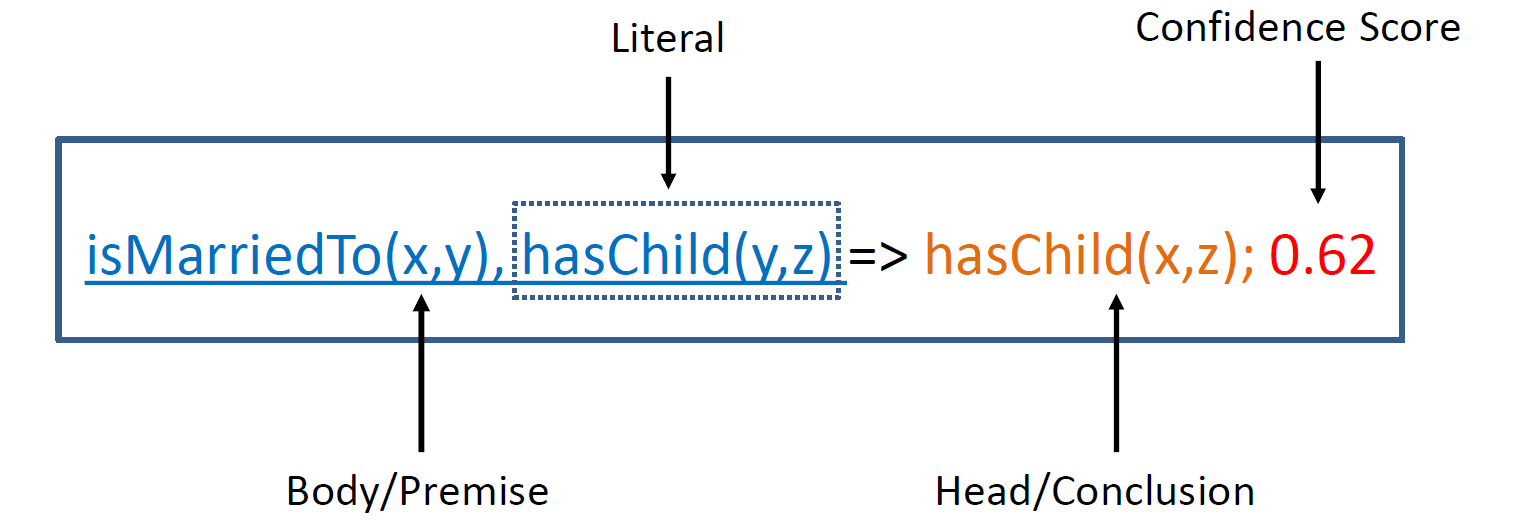}
	\caption{An example rule.} \label{frule}
\end{figure}

Here we further explain the horn-clause logical rules used in our system, which are
pre-learned by previous work \cite{chen2016ontological}.
A Horn clause is a disjunction of literals.
An example horn-clause rule is shown in Figure~\ref{frule}.

In Figure~\ref{frule}, the premise $isMarriedTo(x,y) \land hasChild(y,z)$ is
called body and the conclusion $hasChild(x,z)$ is called head.
Each rule has a confidence score indicating its validness.
There are two kinds of rules used in our system divided by the numbers of literals
in their bodies, length-1 rules (e.g. $diedIn(x,y) \implies wasBornIn(x, y)$) and
length-2 rules (e.g. $isMarriedTo(x,y) \newline \land hasChild(y,z) \implies hasChild(x,z)$).
Also premises can contain reverse relations, such as \textit{hasChild$^-$(x, y)}

We eliminated noisy and incorrect rules learned by previous work \cite{chen2016ontological}
and discarded rules with very low confidence and support.
As discussed earlier, we expand these rules with two types of literals/edges (KB or QA)
to generate paths in the multimodal knowledge graphs.
The reason of using WebQA to create new edges is, if we only use existing facts
in knowledge bases, our system would have very poor KBC performance,
because these knowledge bases are highly incomplete.

\subsection{Multimodal Knowledge Graph Generation}

For each KBC query, our system generates a multimodal knowledge graph on-the-fly.
Our system first retrieves the set of rules for each query relation, then loads existing
premise facts from knowledge bases and extracts new premise facts from the Web using
WebQA to create paths from query subject entities to candidate answers.
A KBC query usually has only a few true answers (no more than 2 true answers in most cases).
To reduce runtime and avoid generating false answers, we follow the query-driven scheme
to create paths.

\subsubsection{Query-Driven optimization}

Loading existing facts from knowledge bases is very fast, because we store all
the existing facts of knowledge bases in a database.
The bottleneck of MKG generation is running WebQA tasks, especially for length-2 rules.
Each WebQA task takes a few seconds, as shown in our previous work \cite{arxivqa,peng2023web}.

For a length-2 rule, our system first extracts intermediate values using the first literal,
then generates candidate answers with intermediate values using the second literal.
If there are $m$ intermediate values from the first literal and $n$ answers
in average for each intermediate value, then we get $m \times n$ candidate answers
in total, of which most are wrong answers.
In this case, our system would issue $m + 1$ WebQA queries for each rule,
which is very time-consuming.
Although we use multi-threading to parallelize the MKG generation process,
issuing all the available WebQA queries is not efficient.
Another disadvantage is low-confidence results from the first step would
generate many incorrect candidate answers at the second step.

In order to improve KBC performance and system efficiency, we use two parameters
to filter the results generated by WebQA for the first literal, the confidence threshol $t$ and the size limit $k$. 
If the confidence score of an intermediate value is smaller than the threshold $t$, we discard it.
And we only pass at most top $k$ intermediate values ranked by confidence scores to the next step.
We learned the best parameters by empirically running experiments with different
sets of parameters.

\subsection{Multimodal Path Fusion Algorithm}

To rank candidate answers for a KBC query, we calculate the scores of candidate
answers based on paths in the multimodal knowledge graph.
For each candidate answer, we retrieve the multimodal paths linking the query entity
to it and calculate the ranking score of it based on all the paths available.
Each path has a confidence score $score\_p$ and weight $weight\_p$.
Path confidence score $score\_p$ measures the confidence of the candidate answer generated by this path.
Path weight $weight\_p$ measures the importance of this path compared to other paths.
The equation to calculate the score of a candidate answer is shown below:
\begin{equation}
\label{pe}
score_{ans} = \sum_{p} score_p \times weight_p \newline
\end{equation}

The confidence score of a path is determined by the confidence scores of its edges.
If the path only has one edge, $score_p = score_{e_0}$.
If the path has two edges, $score_p = score_{e_0} \times score_{e_1}$.
If an edge exists in knowledge bases, its confidence score is 1.0.
If an edge is extracted from the Web, its confidence score is generated by WebQA, which is
illustrated in details in our previous work \cite{arxivqa,peng2023web}.
How to learn path weights is described below.

\subsubsection{Learning Path Weights}

Path weights are used to measure the usefulness of paths towards ranking candidate answers.
We learn the path weights on training datasets through two approaches.

The first approach is to calculate the frequencies of paths appearing with correct answers
on training datasets.
Path frequency indicates the relevance between paths and correct answers.
Path frequency equals the number of times a path cooccurs with the correct answers
divided by total number of appearances of this path in training datasets.

The second approach is to use logistic regression to learn the path importance
on training datasets.
We trained logistic regression classifiers with L2 regularization on path feature vectors to determine a candidate answer is true or false.
The path feature vectors are occurrence vectors of distinct paths for each candidate answer.
The feature importance scores generated from logistic regression are used as path weights.
The training datasets are highly imbalanced with much more negative samples than
positive samples, so we used resampling to balance the datasets.

\subsection{Baseline}

\begin{table*}[ht]
  \caption{KBC performance of WebQA, rule inference, ensemble fusion and multimodal path Fusion
   (measured by MAP).}
  \label{overall}
  \small
  \begin{tabular}{|c|c|c|c|c|c|}
    \hline
     & \textbf{WebQA} & \textbf{Rule Inference} & \textbf{Ensemble Fusion} & \textbf{MPF-Frequency} & \textbf{MPF-Importance} \\
    \hline
    wasBornIn & 0.70 & 0.11 & 0.74 & 0.79 & \textbf{0.82} \\ \hline
    hasChild & 0.24 & 0.22 & 0.33 & 0.36 & \textbf{0.40} \\   \hline
    isCitizenOf & 0.41 & 0.08 & 0.46 & 0.56 & \textbf{0.68} \\ \hline
    isMarriedTo & 0.51 & 0.11 & 0.50 & 0.54 & \textbf{0.59} \\ \hline
  \end{tabular}
\end{table*}

Since there is no similar prior work on combining question answering and rule inference
for knowledge base completion, we also implement a baseline system using simple
ensemble fusion to compare with our multimodal path fusion algorithm on KBC performance.
The baseline system combines confidence scores from rule inference and WebQA through
a few ensemble approaches.
WebQA generates confidence scores for each candidate answer, which is further explained
in our previous work \cite{arxivqa,peng2023web}.

\subsubsection{Rule Inference}
The facts in the knowledge base are stored as triplets in database tables, so rule
inference is conducted by executing corresponding SQL queries in database.
Because each relation has multiple rules, SQL queries for all these rules are run
in parallel using multi-threading, which is very fast.
Since knowledge bases are highly incomplete, rule inference using only
existing facts inside knowledge bases fails to infer missing facts in many cases.

If a fact can be inferred by more than one rule, we must combine the confidence scores
from multiple rules to determine the confidence score of this fact.
Otherwise, we can directly use the confidence score of the only rule as the
confidence score of the fact.
We tested a few approaches, including the maximum approach (choosing the highest score
of all inferred rules), the sum approach (adding all scores together) and logistic regression
(based on features such as average confidence score and total number of rules
by which the fact is inferred).
The sum approach has the best empirical performance through experiments.
Notice here the final confidence score of an inferred fact could exceed 1.0.

\subsubsection{Ensemble Fusion}

Both WebQA and rule inference produce a list of ranked candidate answers with confidence scores.
Then the baseline system combines their results together through ensemble fusion.
Experimental results in Section 4 demonstrate the baseline system can achieve better
performance than WebQA and rule inference.

We used a few ensemble fusion approaches to combine results from WebQA and rule inference.
Let's say for a candidate answer $ans$, WebQA produces confidence score $score_q$
and rule inference produces confidence score $score_r$.
The ensemble approaches are shown below.
\begin{itemize}
	\item Linear rule. We use linear weights to combine confidence scores of
	candidate answers of WebQA and rule inference. The final confidence score
	$score_f$ for $ans$ after linear rule fusion is
	$score_f = \lambda \times score_q + (1 - \lambda) \times score_r$.
	The linear weight $\lambda$ is calculated based on the performance of WebQA and rule inference
	on training datasets.
	\item Maximum rule. We choose the larger score between $score_q$ and $score_r$
	as the final score for $ans$, which means $score_f = max(score_q, score_r)$.
	\item Sum rule. We directly add the confidence scores of $ans$ from WebQA and rule infernce,
	so $score_f = score_q + score_r$.
	\item Logistic regression. We use $score_q$ and $score_r$ as features for $ans$
	and use a trained logistic regression classifier to classify the feature vector
	for $ans$ and produce the probability score as $score_f$.
\end{itemize}
We then rank the candidate answers by their final confidence scores.
Empirical results demonstrate sum rule works better than the other approaches.

\section{Experimental Results}

In this section, we demonstrate the effectiveness and efficiency of our system
through extensive experiments.
Batch-oriented KBC systems use large datasets for extraction, which do not
suit for our query-driven system.
Other on-the-fly systems do not provide datasets or standard benchmarks.
Since there are no standard benchmark datasets publicly available for query-driven
knowledge base completion, we build our own benchmark datasets for evaluation.
The datasets and evaluation metrics in this paper are similar to those in our previous work \cite{arxivqa,peng2023web}.

For KBC performance, we evaluate the quality of candidate answer rankings using
mean average precision (MAP).
For a KBC query, the average precision is defined as
$AP = (\sum_{k=1}^{n} p(k) \times r(k)) / n $,
where $k$ is the rank in the sequence of candidate answers,
$n$ is the number of candidate answers, $p(k)$ is the precision at cut-off $k$
in the ranked list and $r(k)$ is the change in recall from candidate answers $k-1$ to $k$.
Averaging over all queries yields the mean average precision (MAP).

\subsection{Datasets}

We choose Yago as the knowledge base for its popularity in research community,
its rich ontology and large amount of facts.
We choose four relations (wasBornIn, isMarriedTo, hasChild, isCitizenOf), which
are popular relations frequently studied in previous work.

Yago \cite{yago,suchanek2007yago} is a huge semantic knowledge base, derived from
Wikipedia, WordNet and GeoNames.
Currently, Yago has knowledge of more than 10 million entities
(like persons, organizations, cities, etc.) and contains more than 120 million
facts about these entities.
The whole Yago knowledge base can be downloaded from Yago website
\footnote{\url{http://www.mpi-inf.mpg.de/departments/databases-and-information-systems/research/yago-naga/yago/downloads/}.}.
To collect training and testing data, we make the local closed-world assumption,
which assumes if Yago has a non-empty set of objects $O$ for a given subject-relation pair,
then $O$ contains all the ground-truth objects for this subject-relation pair.

For each relation, we randomly sampled 500 queries (subjects and corresponding objects) from
Yago as training datasets to train the path weights for the multimodal path fusion algorithm and 100 queries for testing the KBC performance of different approaches.

\subsection{KBC Performance}

Here we present the overall performance of the multimodal path fusion (MPF) algorithm,
comparing with ensemble fusion, WebQA and rule inference.
Then we explain the runtime efficiency of our system and how we provide fast responses
to queries on-the-fly.

The KBC performance results are shown in Table~\ref{overall}.
For the experiment setup, WebQA is configured with selected question templates and snippet filtering with top 30 snippets, as shown in our previous work \cite{arxivqa,peng2023web}.
And the sum rule approach is used to combine WebQA and rule inference in ensemble fusion.
MPF-Frequency refers to a variant of the MPF algorithm using path frequency values as path weights,
while MPF-Importance refers to using path importance scores as path weights.
The details of how to generate these two types of path weights can be found in Section 3.3.1.

\subsubsection{WebQA and Rule Inference}

WebQA generally works much better than rule inference, especially for three relations (536\% higher MAP for wasBornIn,
413\% higher MAP for isCitizenOf and 364\% higher MAP for isMarriedTo).
The most important reason is that knowledge bases are highly incomplete. 
Recall is very low for rule inference, although the rules themselves are relatively good in terms of precision.
And WebQA can extract a lot of useful facts about entities based on the vast amount of information available on the Web.

\subsubsection{Ensemble Fusion}
Ensemble fusion shows reasonable improvements over WebQA and rule inference for three relations wasBornIn, hasChild and isCitizenOf.
Compared with the better one of WebQA and rule inference, ensemble fusion achieves 6\% higher MAP for wasBornIn, 38\% higher MAP for hasChild and 12\% higher MAP for isCitizenOf.
The results show that ensemble fusion can capture the complementary relation between WebQA and rule inference in most cases.
For the only exception relation isMarriedTo, the performance loss of 0.01 drop in MAP is actually very small.

\subsubsection{Multimodal Path Fusion}

The two MPF algorithms can achieve much higher performance than WebQA, rule inference and baseline ensemble fusion for all four relations.
The reasons are twofold: first MPF can discover more relevant facts by launching WebQA for length-1 and length-2 rules; second, MPF can use multimodal paths to capture the semantic information in multimodal knowledge graphs.

Between the two MPF variants, path importance outperforms path frequency, 
since it learns better path weights using machine learning, while path frequency is a simple rule-based approach.
Compared to ensemble fusion, the MPF-Importance approach achives 11\% higher MAP for relation wasBornIn, 21\% higher MAP for relation hasChild, 48\% higher MAP for relation isCitizenOf and 18\% higher MAP for relation isMarriedTo.

In conclusion, our multimodal path fusion algorithm can achieve much better KBC performance compared to WebQA, rule inference and simple ensemble fusion, by fusing question answering and rule inference in a very sophisticated way.

\subsection{Efficiency}

To evaluate the efficiency of our system, the experiments were run on a single
machine with a 3.1GHZ four-core CPU and 4GB memory.
The system runtime varies with multiple environment factors such as network congestion and server speed.
So we calculated average runtime through extensive experiments with multiple queries.

In the multimodal path fusion algorithm, multimodal path generation is running in parallel.
Looking up facts inside knowledge bases using SQL queries is very fast.
The bottleneck of path generation is the WebQA operations, because they involve a lot of web searches and server inquiries.
In order to improve system efficiency, we use query-driven filtering to avoid running too
many WebQA queries, which is explained in Section 3.2.1.

Since we only have 1-step and 2-step rules, the runtime of the our system is roughly about
2 times of a single WebQA operation because of parallel path generation.
The efficiency of our WebQA system is explained in our paper \cite{arxivqa,peng2023web}. 
The overall runtime of our system is about 6 seconds in average for each KBC query,
which is very fast considering the fact that we need to issue many WebQA queries,
which launch a lot of web searches and server requests.

\section{Conclusions}

In this paper, we propose an effective and efficient KBC system by
combining rule inference and web-based question answering.
Our system fuses both unstructured data from the Web and structured data from
knowledge bases in depth.
We generate a multimodal knowledge graph for each query based on question answering
and rule inference.
We propose a multimodal path fusion algorithm to effectively rank
candidate answers using the multimodal knowledge graph.
To improve efficiency, we employ query-driven techniques to reduce the
runtime on-the-fly and provide fast responses to user queries.
Extensive experiments have been conducted to demonstrate the effectiveness and
efficiency of our system.

There are several directions for the future work of our project.
First, we can extend our system to full-scale completion on much larger knowledge bases,
such as Freebase.
Building a scalable system that can handle millions of entities and billions of facts
is an important problem worth studying.
Second, we can explore the direction of iteratively updating knowledge bases using our system.


\bibliographystyle{ACM-Reference-Format}
\bibliography{reference}

\end{document}